\documentclass[conference]{IEEEtran}
\IEEEoverridecommandlockouts

\usepackage{cite,comment}
\usepackage{mathtools}
\usepackage{hyperref}
\usepackage{graphicx}
\usepackage{enumitem}
\usepackage{tikz}
\usepackage{bm}
\usepackage{amsmath}
\usepackage{algorithm}
\usepackage{algorithmic}
\usepackage{amsmath,amssymb,amsthm}
\usepackage{array}
\usepackage{relsize}
\usepackage{subcaption}
\usepackage{mathtools, nccmath}
\usepackage[labelformat=simple]{subcaption}
\usepackage{derivative}

\usepackage[top    = 0.80in,
bottom = 1.05in,
left   = 0.80in,
right  = 0.80in]{geometry}

\newcolumntype{P}[1]{>{\centering\arraybackslash}p{#1}}
\DeclareMathOperator{\C}{\mathbb{C}}
\DeclareMathOperator{\E}{\mathbb{E}}
\DeclareMathOperator{\R}{\mathbb{R}}
\DeclareMathOperator{\N}{\mathcal{N}}

\DeclareMathOperator{\CG}{\mathcal{C}}

\DeclareMathOperator*{\argmin}{arg\,min}
\DeclarePairedDelimiter\abs{\lvert}{\rvert}
\DeclareCaptionSubType * [alph]{table}

\newcommand{\A}{\mathbf{A}}

\newcommand{\h}{\mathbf{H}}
\newcommand{\f}{\mathbf{F}}

\newcommand{\I}{\mathbf{I}}

\newcommand{\fs}{\mathbf{f}}

\newcommand{\js}{{\rm j}}

\newcommand{\hs}{\mathbf{h}}

\newcommand{\sn}{\mathsf{s}}

\newcommand{\ysn}{\mathsf{y}}
\newcommand{\ys}{\bm{\mathsf{y}}}
\newcommand{\s}{\bm{\mathsf{s}}}
\newcommand{\ws}{\bm{\mathsf{w}}}
\newcommand{\wsn}{\mathsf{w}}

\newcommand{\hr}{\mathsf{H}}
\newcommand{\tr}{\mathsf{T}}

\newcommand{\bc}{\begin{center}}
\newcommand{\ec}{\end{center}}
\newcommand{\norm}[1]{\left\lVert#1\right\rVert}
\newcommand{\ds}{\displaystyle}
\newcommand{\uprightsubscript}[1]{_{\textnormal{#1}}}
\begingroup\lccode`~=`?\lowercase{\endgroup\let~}\uprightsubscript
\AtBeginDocument{\mathcode`?="8000 }

\newcommand{\tn}[1]{\textnormal{#1}}

\renewcommand{\baselinestretch}{1.0}
\theoremstyle{remark}

\definecolor{dg}{RGB}{255,0,0}

\DeclareRobustCommand{\IEEEauthorrefmark}[1]{\smash{\textsuperscript{\footnotesize #1}}} 
\begin{document}

\title{Electromagnetically-Consistent Modeling and Optimization of Mutual Coupling in RIS-Assisted Multi-User MIMO Communication Systems
}
\author{\IEEEauthorblockN{Dilki Wijekoon\IEEEauthorrefmark{1}, Amine Mezghani\IEEEauthorrefmark{1}, George C. Alexandropoulos\IEEEauthorrefmark{2}, and Ekram Hossain\IEEEauthorrefmark{1}} 
	\IEEEauthorblockA{\IEEEauthorrefmark{1}Department of Electrical and Computer Engineering, University of Manitoba, Winnipeg, Canada 
}
\IEEEauthorblockA{\IEEEauthorrefmark{2}Department of Informatics and Telecommunications, National and Kapodistrian University of Athens, Greece
 }
 \IEEEauthorblockA{Emails: wijekood@myumanitoba.ca, \{amine.mezghani, ekram.hossain\}@umanitoba.ca, alexandg@di.uoa.gr
 }
}

\maketitle

\begin{abstract}
Mutual Coupling (MC) is an unavoidable feature in Reconfigurable Intelligent Surfaces (RISs) with sub-wavelength inter-element spacing. Its inherent presence naturally leads to non-local RIS structures, which can be efficiently described via non-diagonal phase shift matrices. In this paper, we focus on optimizing MC in RIS-assisted multi-user MIMO wireless communication systems. We particularly formulate a novel problem to jointly optimize active and passive beamforming as well as MC in a physically consistent manner. To characterize MC, we deploy scattering parameters and propose a novel approach to optimize them through an offline optimization method, rather than optimizing MC on the fly. Our numerical results showcase that the system performance increases with the proposed MC optimization, and this improvement is achievable without the need for optimizing MC on-the-fly, which can be rather cumbersome.
\end{abstract}
\vspace{3pt}

\section{Introduction} 
Reconfigurable Intelligent Surfaces (RISs) constitute a technology that has the potential to revolutionize the performance, coverage, security, and efficiency of wireless communication systems \cite{ABoI_EURASIP,RISsurvey2023,RIS_ISAC_SPM}. Their operation involves adjusting the phases of impinging signals, providing dynamically programmable generalized reflections. Joint active and passive (reflective) beamforming optimization enhances the signal quality and strength for intended users, while simultaneously reducing interference caused by unintended users \cite{huang2019reconfigurable,9148961,9474428}.
In the context of RISs, choosing a convenient phase shift matrix is critical for their configuration. The recent studies \cite{9913356,9737373} present RIS models employing non-diagonal phase shift matrices, also known as non-local RIS structures \cite{10052027}, which outperform conventional RIS models. However, these structures require extra hardware with physical connections among RIS elements, resulting in increased wiring complexity and control overhead. Fortunately, non-local RIS structures can also be naturally attainable when Mutual Coupling (MC) between elements exists.

MC is an unavoidable effect occurring among adjacent RIS elements due to their Electromagnetic (EM) interactions~\cite{PhysFad}. When the RIS elements are placed in close vicinity to each other, especially with sub-wavelength spacing, MC becomes strong and influences their EM behavior. Several works consider the MC impact and its consequences on RIS-assisted wireless systems \cite{PhysFad,9360851,9525465,9319694,AbrardoAndrea2023AaOo,10096563,rabault2023tacit}, with some of them using the impedance $\mathrm{Z}$-parameters for its characterization~\cite{9360851,9525465,9319694}. The latter works assume minimum scattering assumptions to derive the impedance matrices, which are, however, impractical. On the other hand, the works~\cite{10096563,AbrardoAndrea2023AaOo} focus on the scattering $\mathrm{S}$-parameters to characterize MC. Irrespective of the modeling approach, all latter works confirm that, when MC is present and taken under consideration during the RIS optimization process, the system performance improves. However, none of these works explicitly deals with efficiently optimizing MC. 

In this paper, we propose a novel approach to simultaneously optimize active and passive beamforming in RIS-assisted multi-user Multiple-Input Multiple-Output (MIMO) systems, which includes MC optimization within a physically-consistent framework, based on the $\mathrm{S}$-parameters, resulting in a non-local RIS structure. The contributions of this work are summarized as follows:
\begin{itemize}
    \item We formulate a novel joint optimization problem for active and passive beamforming as well as MC for RIS-aided multi-user downlink transmission based on a physically-consistent RIS model. We particularly optimize MC at the RIS through an offline approach for an ensemble of wireless channels, rather than optimizing it dynamically on the fly for each channel realization.
    \item The offline MC optimization exhibits a complex nested structure within the overall optimization problem, involving both outer and inner optimization problems. The inner problem deals with the optimization of the RIS phase configuration and the Base Station (BS) active precoding, while the outer problem involves optimizing MC based on the scattering $\mathrm{S}$-parameters. The inner problem is addressed using the method described in \cite{10096563}, while the outer problem is solved via a combination of the projected gradient descent method and the method of Lagrange multipliers.
    \item Our simulation results showcase that the system performance is further enhanced through joint beamforming and MC optimization. This improvement is achieved even with the proposed offline process for optimizing the MC at the RIS, emphasizing the effectiveness beyond real-time MC adjustments.
\end{itemize}
The rest of the paper is organized as follows. Section II outlines the system model, assumptions, and optimization problem formulation. Section III describes the proposed solution approach, while Section IV describes the simulation results. Finally, the paper is concluded in Section V.

\textbf{Notations}: Scalar variables are denoted by lowercase letters, while vectors and matrices are represented by small and capital boldface letters (e.g., $\s$ and $\mathbf{S}$). The transpose, conjugate, and conjugate transpose are denoted by $(.)^\tr$, $(.)^*$, and $(.)^\hr$, respectively. Notations $\|.\|?2$ and $\|.\|?F$ stand for the Euclidean and Frobenius norms, respectively, whereas the trace and rank of a matrix are represented with $\tn{Tr}(\A)$ and $\tn{Rank}(\A)$, respectively.
The expectation operator is indicated by $\E\lbrace.\rbrace$. The term $\tn{Diag}(\bm a)$ refers to a diagonal matrix whose diagonal elements are the components of $\bm a$. $\otimes$ is the Kronecker product, whereas $\textbf{I}_{M}$ and $\textbf{0}_{M}$ are the $M \times M$ identity and all-zeros matrices, respectively.
\begin{figure}[!t]
\includegraphics[scale=0.30]{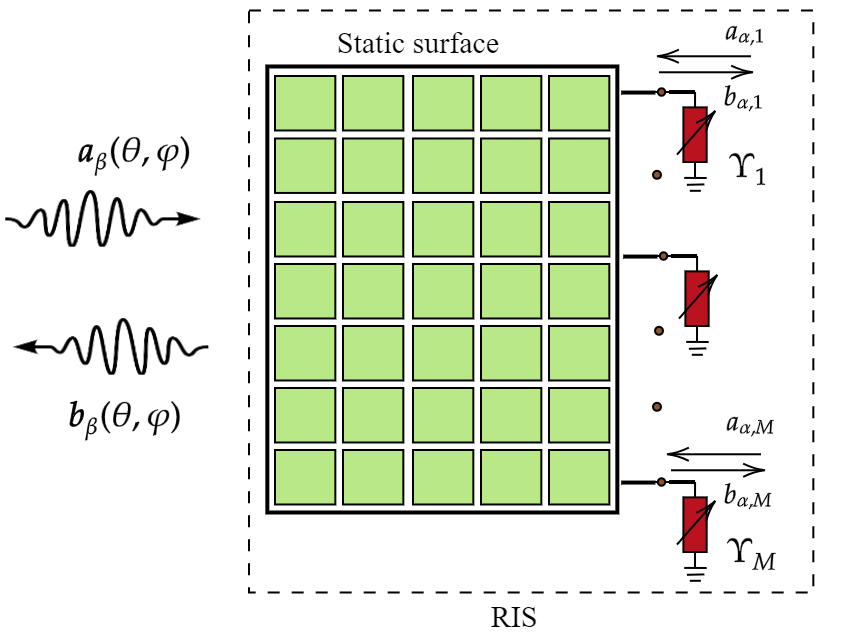}
\caption{Transmit/receive space- and port-side RIS patterns.}
\label{fig:1}\vspace{-5mm}
\end{figure}

\section{System Model and Problem Formulation}
\subsection{Physically-Consistent RIS Modeling}
The EM properties of an RIS are explained by its transmit/receive space- and port-side scattering patterns \cite{10052027}. According to Fig.~\ref{fig:1}, the radiation pattern of the space side is represented by the incoming and outgoing wave phasor vectors $\boldsymbol{a_{\beta}}(\theta,\varphi)$ and $\boldsymbol{b_{\beta}}(\theta,\varphi)$, respectively, with respect to the azimuth and elevation angles $\theta$ and $\varphi$ on the horizontal and vertical axes, respectively. 
Moreover, the forward and backward traveling wave phasors of each element corresponding to the port-side scattering are denoted by $a_{\alpha,m}$ and $b_{\alpha,m}$, respectively, where $m$ (with $m=1,2,\dots,M$) denotes the RIS element index. The mathematical representations of space- and port side-scattering are related as follows:  
\begin{equation}
\small
\label{eqn:sc}
\underbrace{\left[ \begin{array}{c}  \boldsymbol{b_{\beta}}(\theta,\varphi)\\ \boldsymbol{b_{\alpha}} \end{array} \right]}_{\triangleq\mathlarger{\boldsymbol{b}}} = \underbrace{\begin{bmatrix} \mathbf{S_{\beta\beta}} & \mathbf{S_{\alpha\beta}}\\ \mathbf{S_{\beta\alpha}} & \mathbf{S_{\alpha\alpha}} \end{bmatrix}}_{\triangleq\mathlarger{\mathbf{S}}} \underbrace{\left[ \begin{array}{c}  \boldsymbol{a_{\beta}}(\theta,\varphi)\\ \boldsymbol{a_{\alpha}}  \end{array} \right]}_{\triangleq\mathlarger{\boldsymbol{a}}}.
\tag{1}
\end{equation}
\normalsize
In this expression, the matrix $\mathbf{S}$ denotes the total scattering matrix of the RIS. $\mathbf{S_{\alpha\alpha}}$ and $\mathbf{S_{\beta\beta}}$ are called ``multi-port" and ``wave" scattering matrices, respectively. In addition, the matrices $\mathbf{S_{\alpha\beta}}$ and $\mathbf{S_{\beta\alpha}}$ imply transmit and receive radiation patterns indicating MC. The terms $\boldsymbol{a_{\alpha}}\triangleq[a_{\alpha,1}, a_{\alpha,2},...,a_{\alpha,M}]^\tr$ and $\boldsymbol{b_{\alpha}}\triangleq[b_{\alpha,1}, b_{\alpha,2},...,b_{\alpha,M}]^\tr$ include respectively the incoming and outgoing wave vectors at all RIS port sides. 

In RIS deployments, each port side is terminated by a load. The interaction between incoming and outgoing signals at each port side is characterized by the ``load" scattering matrix $\mathbf{\Upsilon}$, which is represented as follows: 
\begin{equation}
\small
\label{eqn:7-n11}
\boldsymbol{b_{\alpha}}=\mathbf{\Upsilon}\boldsymbol{a_{\alpha}}.
\tag{2}
\end{equation}
\normalsize
Expanding this expression yields the following equations:
\begin{equation}
\small
\label{eqn:7-n1}
\boldsymbol{b_{\beta}}(\theta,\varphi)=\mathbf{S_{\beta\beta}}\boldsymbol{a_{\beta}}(\theta,\varphi)+\mathbf{S_{\alpha\beta}}\boldsymbol{a_{\alpha}},
\tag{3}
\end{equation}
\normalsize
\begin{equation}
\small
\label{eqn:7-n2}
\boldsymbol{b_{\alpha}}=\mathbf{S_{\beta\alpha}}\boldsymbol{a_{\beta}}(\theta,\varphi)+\mathbf{S_{\alpha\alpha}}\boldsymbol{a_{\alpha}}.
\tag{4}
\end{equation}
\normalsize
By utilizing (\ref{eqn:7-n11}), (\ref{eqn:7-n1}), and (\ref{eqn:7-n2}), the total outgoing wave phasor $\boldsymbol{b_{\beta}}(\theta,\varphi)$ at the space-side scattering pattern is given by
\begin{align*}
\small
\label{eqn:e1} 
\boldsymbol{b_{\beta}}(\theta, \varphi)=  \underbrace{\mathbf{S_{\beta \alpha}}\left(\mathbf{\Upsilon}^{-1}-\mathbf{S_{\alpha\alpha}}\right)^{-1}\mathbf{S_{\alpha \beta}}\boldsymbol{a_{\beta}}(\theta,\varphi) }_{\text {Adaptive scattering }}+\underbrace{\mathbf{S_{\beta \beta}}\boldsymbol{a_{\beta}}(\theta,\varphi)}_{\text {Residual scattering}}. \tag {5} 
\end{align*}   
\normalsize
The term $\left(\mathbf{\Upsilon}^{-1}-\mathbf{S}_{\alpha\alpha}\right)^{-1}$ can be expanded as a summation of higher order terms, as shown in the following expression known as ``Neumann series approximation''~\cite{10096563,rabault2023tacit}:
\begin{equation}
\small
\label{eqn:e2}
(\mathbf{\Upsilon}^{-1}-\mathbf{S}_{\alpha\alpha})^{-1}=\ds\sum_{l=0}^\infty(\mathbf{\Upsilon}\mathbf{S}_{\alpha\alpha})^{l}\mathbf{\Upsilon}=\mathbf{\Upsilon}+\mathbf{\Upsilon}\mathbf{S}_{\alpha\alpha}\mathbf{\Upsilon}+\cdots. \tag{6}
\end{equation}
\normalsize
Conventional models utilize a simplified version of (\ref{eqn:e1}) that ignores residual scattering and assume that $\mathbf{S}_{\alpha\beta}=\mathbf{S}_{\beta\alpha}=\mathbf{I}_M$ and absence of multiple reflections, i.e., $\mathbf{S}_{\alpha\alpha}=\mathbf{0}_M$. In this paper, we incorporate MC and multiple reflections \cite{PhysFad,10096563,rabault2023tacit}, and optimize the associated scattering parameters $\mathbf{S}_{\alpha\beta}$, $\mathbf{S}_{\beta\alpha}$, and $\mathbf{S}_{\alpha\alpha}$ for a communications' objective. For simplicity, we also ignore the residual scattering part in (\ref{eqn:e1}). 

\subsection{System Model}
Consider a multi-user downlink MIMO system comprising a single BS equipped with $N$ antennas, $K<N$ single-antenna users, and a single RIS with $M>K$ elements. The direct channel between the BS and any of the users is assumed to be blocked. The matrix $\h?{r-u} \triangleq [\hs_{\tn{r-u},1}, \hs_{\tn{r-u},2},\ldots, \hs_{\tn{r-u},K}]$ includes all the RIS-user channel gains, with $\hs_{\tn{r-u},k} \in \C^{M}$ representing each RIS-$k$-th user link ($k=1,2,\ldots,K$), while the matrix $\h?{b-r} \in \C^{M\times N}$ corresponds to the BS-RIS channel gains and satisfies the condition: $\tn{Rank}(\h?{b-r})\geq K$.

It follows from the previous subsection that the effective RIS phase shift matrix is given by $\boldsymbol{\Phi}\triangleq\left(\mathbf{\Upsilon}^{-1}-\mathbf{S}_{\alpha\alpha}\right)^{-1}$, where $\mathbf{\Upsilon}=\tn{Diag}(\bm\upsilon) \in \C^{M \times M}$ with $\bm\upsilon \in \C^M$ including the tunable phase shifts applied to the ports: $\upsilon_m=e^{\js\theta_m}$ ($m=1,2,\ldots,M$) with $\abs*{\upsilon_m}=1$ and $\theta_m\in[0,2\pi]$. Putting all above together, the baseband received signals at all $K$ users can be mathematically expressed as follows:
\begin{equation}
\label{eqn:eq3}
\ys\triangleq\rho\h?{r-u}^\hr\mathbf{S}_{\beta\alpha}{(\mathbf{\Upsilon}^{-1}-\mathbf{S}_{\alpha\alpha})^{-1}}\mathbf{S}_{\alpha\beta}\h?{b-r}\f\s +\ \rho\ws, \tag {7} 
\end{equation}
where $\ys=[\ysn_1, \ysn_2, \ldots, \ysn_K]^\tr$ and $\s\triangleq[\sn_1,\sn_2,\ldots,\sn_K]^\tr$ represents the transmit symbol vector such that $\sn_k \sim \CG\N(0,1)$ $\forall k$. The receiver scaling factor $\rho \in \R$ and is assumed to be common for all users. The vector $\ws$ denotes the Additive White Gaussian Noise (AWGN) containing the noise components $\wsn_k \sim \CG\N(0,\sigma?w^2)$. Matrix $\f\triangleq[\fs_1, \fs_2, \ldots, \fs_K]$ with $\fs_k \in \C^{N}$ represents the transmit beamforming matrix for which it holds
$\E_{\s}\big\{\|\f\s\|^2\big\}= P$ with $P$ being the total transmit power. We define the end-to-end RIS-parametrized multi-user channel as follows:
\begin{equation}
\label{eqn:eq5}
\h^\hr\triangleq\h?{r-u}^\hr\mathbf{S}_{\beta\alpha}{(\mathbf{\Upsilon}^{-1}-\mathbf{S}_{\alpha\alpha})^{-1}}\mathbf{S}_{\alpha\beta}\h?{b-r}. \tag {8}
\end{equation}
It is noted that, in \cite{10096563}, it was assumed for the radiation patterns that $\mathbf{S}_{\beta\alpha}=\mathbf{S}_{\alpha\beta}=\mathbf{I}_M$. 

We now define the Minimum Mean Squared Error (MMSE) criterion~\cite{9474428,10096563} 
which will serve as our system's design optimization objective. The connection between the achievable sum-rate performance, $C$, and the MMSE of each of the $K$ users, $\tn{MMSE}_k$, can be expressed as follows:
\begin{equation}
\label{eqn:eq4}
C \triangleq \ \ds\sum_{k=1}^K\log_2\left(\frac{1}{\tn{MMSE}_k}\right) \ = \ \log_2\left(\ds\prod_{k=1}^K\frac{1}{\tn{MMSE}_k}\right), \tag {9} 
\end{equation}
with the total MSE for all $K$ users given by:
\begin{equation}
\label{eqn:eq5_1}
\ds\sum_{k=1}^K\E_{\ysn_k,\sn_k}\left\lbrace\abs*{\ysn_k-\sn_k}^2\right\rbrace \ = \E_{\ys,\s}\left\lbrace\norm{\ys-\s}_2^2\right\rbrace, \tag {10}
\end{equation} 
with $\E_{\ysn_k,\sn_k}\left\lbrace\abs*{\ysn_k-\sn_k}^2\right\rbrace$ indicating the MSE of the received symbol for each $k$-th user. 

\subsection{Problem Formulation}\label{subsec:Problem}
Finding the scattering parameters for each channel that optimize the sum rate in~\eqref{eqn:eq4} on the fly (i.e., in the online phase) is impractical, since this will require fabricating the RIS accordingly. Therefore, in this paper, we focus on optimizing the scattering matrices $\mathbf{S_{\alpha\alpha}}$, $\mathbf{S_{\alpha\beta}}$, and $\mathbf{S_{\beta\alpha}}$ offline for a particular class of channels, and then, optimize $\f$ and $\mathbf{\Upsilon}$ online (i.e., for each channel realization). In particular, we formulate the following optimization problem:
\begin{subequations}
\label{eqn:eq6-}
\begin{align}
\label{eqn:eq6-a}
\tag{11a}
\ds\argmin_{ \mathbf{S_{\alpha\alpha}},  \mathbf{S_{\alpha\beta}}, \mathbf{S_{\beta\alpha}}}  \quad & \mathbb{E}_{\h?{r-u},\h?{b-r}}\left[\ds\argmin_{\rho, \f, {\mathbf\Upsilon}} \E_{\ys,\s}\left\lbrace\|\ys-\s\|_2^2\right\rbrace \right] \\ 
\label{eqn:eq6-b}
\tn{subject to} \quad & \tag{11b}
\E_{\s}\left\lbrace\|\f\s\|_2^2\right\rbrace=P,\\ 
\label{eqn:eq6-c} \tag{11c}
& \upsilon_{im}=0, \quad \forall i \neq m,\\ 
\label{eqn:eq6-d} \tag{11d}
& |\upsilon_{ii}|=1, \quad \forall i=1,2, \ldots, M, \\ 
\label{eqn:eq6-e} \tag{11e} 
& \mathbf{S_{\alpha\alpha}}\mathbf{S}_{\alpha\alpha}^\hr+ \mathbf{S}_{\alpha\beta}\mathbf{S}_{\alpha\beta}^\hr=\I_{M}, \\
\label{eqn:eq6-e2} \tag{11f}
& \mathbf{S}_{\alpha\beta}=\mathbf{S}_{\beta\alpha}^\tr \textrm{ and } \mathbf{S}_{\alpha\alpha}=\mathbf{S}_{\alpha\alpha}^\tr.
\end{align}  
\end{subequations}
Constraint \eqref{eqn:eq6-b} is the power allocation constraint, whereas \eqref{eqn:eq6-c} and \eqref{eqn:eq6-d} are unimodular constraints related to the RIS phase configuration matrix. Furthermore, the last two constraints stand for the losslessness and reciprocity of the RIS reflective structure. Using (\ref{eqn:eq3}) and (\ref{eqn:eq5}), the latter optimization problem can be rewritten as follows:
\begin{subequations}
\label{eqn:eq7}
\begin{align*}
   \tag{12a}
    \label{eqn:eq7-a}
     \ds\argmin_{ \mathbf{S_{\alpha\alpha}},  \mathbf{S_{\alpha\beta}}, \mathbf{S_{\beta\alpha}}} \,\, & \mathbb{E}_{\h?{r-u},\h?{b-r}} \Bigl\{\ds\argmin_{\rho, \f, {\mathbf\Upsilon}}\norm{\rho\h^\hr\f -\I_K}?F^2 + K\rho^2\sigma?w^2\Bigr\}
    \vspace{5pt} \\
\label{eqn:eq7-b} \tag{12b}
\tn{subject to} \quad & \|\f\|?F^2=P,\\
\label{eqn:eq7-c} \tag{12c}
& \upsilon_{im}=0, \quad \forall i \neq m,\\
\label{eqn:eq7-d} \tag{12d}
& |\upsilon_{ii}|=1, \quad \forall i=1,2, \ldots, M, \\
\label{eqn:eq7-e} \tag{12e} 
& \mathbf{S}_{\alpha\alpha}\mathbf{S}_{\alpha\alpha}^\hr+ \mathbf{S}_{\alpha\beta}\mathbf{S}_{\alpha\beta}^\hr=\I_{M}, \\
\label{eqn:eq7-e2} \tag{12f}
& \mathbf{S}_{\alpha\beta}=\mathbf{S}_{\beta\alpha}^\tr \textrm{ and } \mathbf{S}_{\alpha\alpha}=\mathbf{S}_{\alpha\alpha}^\tr.
\end{align*}  
\end{subequations}
By incorporating the equality $\mathbf{S_{\alpha\beta}}=\mathbf{S}^\tr_{\beta\alpha}$ into the problem's objective function \eqref{eqn:eq7-a} and using the matrix notation $\hat{\h}^\hr\triangleq\h?{r-u}^\hr\mathbf{S}_{\alpha\beta}^\tr{(\mathbf{\Upsilon}^{-1}-\mathbf{S}_{\alpha\alpha})^{-1}}\mathbf{S}_{\alpha\beta}\h?{b-r}$, yields the following reformulation of the considered design problem:
\begin{subequations}
\vspace{-8pt}
\label{eqn:eq8}
\begin{align*}
   \tag{13a}
     \label{eqn:eq8-a}
       \ds\argmin_{ \mathbf{S_{\alpha\alpha}},  \mathbf{S_{\alpha\beta}}} \quad & \mathbb{E}_{\h?{r-u},\h?{b-r}} \Bigl\{\ds\argmin_{\rho, \f, {\mathbf\Upsilon}}\norm{\rho\hat{\h}^\hr\f -\I_K}?F^2 + K\rho^2\sigma?w^2 \Bigr\}
    \\
\label{eqn:eq8-b} \tag{13b}
\tn{subject to} \quad & \|\f\|?F^2=P,\\
\label{eqn:eq8-c} \tag{13c}
& \upsilon_{im}=0, \quad \forall i \neq m,\\
\label{eqn:eq8-d} \tag{13d}
& |\upsilon_{ii}|=1, \quad \forall i=1,2,\ldots, M, \\
\label{eqn:eq8-e} \tag{13e} 
& \mathbf{S}_{\alpha\alpha}\mathbf{S}_{\alpha\alpha}^\hr+ \mathbf{S}_{\alpha\beta}\mathbf{S}_{\alpha\beta}^\hr=\I_{M}, \\
\label{eqn:eq8-f} \tag{13f} 
& \mathbf{S}_{\alpha\alpha}=\mathbf{S}_{\alpha\alpha}^\tr.
\end{align*}  
\end{subequations}
To reduce the complexity of the optimization process, we next express the scattering matrices as $\mathbf{S_{\alpha\alpha}}=\textbf{U}\mathbf{\Sigma_{\alpha\alpha}}\textbf{V}^\hr$ and $\mathbf{S_{\alpha\beta}}=\textbf{U}\mathbf{\Sigma_{\alpha\beta}}\textbf{V}^\hr$, where $\mathbf{\Sigma_{\alpha\alpha}}$ and $\mathbf{\Sigma_{\alpha\beta}}$ are real diagonal matrices whose respective $M$ non-zero elements will be optimized, and the $M\times M$ matrices $\textbf{U}$ and $\textbf{V}$ will be considered as fixed throughout the process. In particular, both $\textbf{U}$ and $\textbf{V}$ are generated as $\mathbf{D}\otimes\mathbf{D}$ with $\mathbf{D}\in\mathbb{C}^{\sqrt{M} \times \sqrt{M}}$ being the Discrete Fourier Transform (DFT) matrix. Using this simplification and the notation $\Tilde{\mathbf{H}}^\hr\triangleq \h?{r-u}^\hr(\textbf{U}\mathbf{\Sigma}_{\alpha\beta}\textbf{V}^\hr)^\tr{(\mathbf{\Upsilon}^{-1}-\textbf{U}\mathbf{\Sigma}_{\alpha\alpha}\textbf{V}^\hr)^{-1}}(\textbf{U}\mathbf{\Sigma}_{\alpha\beta}\textbf{V}^\hr)\h?{b-r}$, the following formulation is deduced: 
\begin{subequations}
\label{eqn:eq9}
\begin{align*}
  \tag{14a}
    \begin{split}
     \label{eqn:eq9-a}
      \ds\argmin_{ \mathbf{\Sigma_{\alpha\alpha}},  \mathbf{\Sigma_{\alpha\beta}}} \quad & \mathbb{E}_{\h?{r-u},\h?{b-r}} \Bigl\{\ds\argmin_{\rho, \f, {\mathbf\Upsilon}}\norm{\rho\Tilde{\mathbf{H}}^\hr\f -\I_K}?F^2 +K\rho^2\sigma?w^2 \Bigr\}
    \end{split} \\
    \vspace{5pt}
\label{eqn:eq9-b_1} \tag{14b}
\tn{subject to} \quad & \|\f\|?F^2=P,\\
\label{eqn:eq9-c} \tag{14c}
& \upsilon_{im}=0, \quad \forall i \neq m,\\
\label{eqn:eq9-d} \tag{14d}
& |\upsilon_{ii}|=1, \quad \forall i=1,2,\ldots, M, \\
\label{eqn:eq9-e} \tag{14e} 
&\mathbf{\Sigma}_{\alpha\alpha}\mathbf{\Sigma}_{\alpha\alpha}^\hr+\mathbf{\Sigma}_{\alpha\beta}\mathbf{\Sigma}_{\alpha\beta}^\hr=\I_{M}, \\
\label{eqn:eq9-f} \tag{14f} 
& \mathbf{\Sigma}_{\alpha\alpha}=\mathbf{\Sigma}_{\alpha\alpha}^\tr.
\end{align*}  
\end{subequations}
\section{Proposed EM-Consistent System Design}
The optimization problem proposed in the previous Section~\ref{subsec:Problem} involves inner and outer minimization sub-problems, and is thus difficult to solve. In this section, we present a decomposition of the overall system design into two sub-problems and deploy an alternating optimization approach. The two sub-problems are the following:
\vspace{15pt}
\begin{enumerate}
\item For given $\mathbf{\Sigma}_{\alpha\alpha}$ and $\mathbf{\Sigma}_{\alpha\beta}$, solve:
\begin{subequations}
\label{eqn:ao-precod-mat_1}
\begin{align*}
\label{eqn:ao-precod-mata} \tag{15a}
\argmin_{\rho, \f,{\mathbf\Upsilon}} \quad & \norm{\rho\Tilde{\mathbf{H}}^\hr\f -\I_K}?F^2 +K\rho^2\sigma?w^2 \\\
\label{eqn:ao-precod-matb}\tag{15b}
\tn{subject to} \quad & \|\f\|?F^2=P,\\ 
\label{eqn:ao_2} \tag{15c}
& \upsilon_{im}=0, \quad \forall i \neq m,\\
\label{eqn:ao_2_d} \tag{15d}
& |\upsilon_{ii}|=1, \quad \forall i=1,2, \ldots, M. \\
\end{align*}
\end{subequations}
\item For given $\rho$, $\f$, and ${\mathbf\Upsilon}$, solve:
\begin{subequations}
\label{eqn:ao-phase-mat_1}
\begin{align*}
\label{eqn:ao-phase-mata} \tag{16a}
\ds\argmin_{ \mathbf{\Sigma}_{\alpha\alpha},  \mathbf{\Sigma}_{\alpha\beta}} \quad & \mathbb{E}_{\h?{r-u},\h?{b-r}} \Bigl\{\norm{\rho\Tilde{\mathbf{H}}^\hr\f -\I_K}?F^2 +K\rho^2\sigma?w^2\Bigr\} \\\
\label{eqn:eq91-b}\tag{16b}
\tn{subject to} \quad & \mathbf{\Sigma}_{\alpha\alpha}\mathbf{\Sigma}_{\alpha\alpha}^\hr+\mathbf{\Sigma}_{\alpha\beta}\mathbf{\Sigma}_{\alpha\beta}^\hr=\I_{M}, \\
\label{eqn:eq91-c} \tag{16c}
& \mathbf{\Sigma}_{\alpha\alpha}=\mathbf{\Sigma}_{\alpha\alpha}^\tr.
\end{align*}
\end{subequations}
\end{enumerate}
The first sub-problem can be efficiently solved via the method described in \cite{10096563}. The solution to the second sub-problem, which gives the optimal scattering matrices, is next obtained via a gradient step and a closed-form projection step for both $\mathbf{\Sigma_{\alpha\alpha}}$ and $\mathbf{\Sigma_{\alpha\beta}}$ matrices. 
Recall from the previous section that we have used the assumptions $\mathbf{S_{\alpha\alpha}}=\textbf{U}\mathbf{\Sigma_{\alpha\alpha}}\textbf{V}^\hr$ and $\mathbf{S_{\alpha\beta}}=\textbf{U}\mathbf{\Sigma_{\alpha\beta}}\textbf{V}^\hr$, where $\mathbf{\Sigma_{\alpha\alpha}}$ and $\mathbf{\Sigma_{\alpha\beta}}$ are real diagonal matrices and $\textbf{U}$ and $\textbf{V}$ are fixed DFT-based matrices. Hence, the optimization of the second sub-problem is conducted with respect to $\mathbf{\Sigma_{\alpha\alpha}}$ and $\mathbf{\Sigma_{\alpha\beta}}$ instead of their corresponding scattering parameters. 

\subsubsection{Projected Gradient Descent}
For each channel realization, we define the following MSE function:
\begin{equation}
\textit{f}(\mathbf{\Sigma_{\alpha\alpha}},\mathbf{\Sigma_{\alpha\beta}})\triangleq  \norm{\rho\Tilde{\mathbf{H}}^\hr\f -\I_K}?F^2 +K\rho^2\sigma?w^2. \tag{17}
\end{equation}
Since the objective function (\ref{eqn:ao-phase-mata}) involves an expectation with respect to $\h?{r-u}$ and $\h?{b-r}$, Monte Carlo sampling can be used to compute the gradients for a set of $Q$ channel samples. To this end, the overall gradient of this objective function with respect to $\mathbf{\Sigma_{\alpha\alpha}}$ is computed as follows:
\begin{equation}\label{eq:G_aa}
G_{\mathbf{\Sigma_{\alpha\alpha}}}\triangleq\frac{1}{Q}\sum_{q=1}^Q \left[\pdv{\textit{f}(\mathbf{\Sigma_{\alpha\alpha}},\mathbf{\Sigma_{\alpha\beta}})}{\mathbf{\Sigma_{\alpha\alpha}}}\right]_\textit{q},\tag{18}
\end{equation}
where the gradient for each $q$-th channel sample is given by:
\begin{align}
\scriptsize
\begin{split}
&\left[\pdv{\textit{f}(\mathbf{\Sigma_{\alpha\alpha}},\mathbf{\Sigma_{\alpha\beta}})}{\mathbf{\Sigma_{\alpha\alpha}}}\right]_\textit{q}=2(\mathbf{\Upsilon}_\textit{q}^{-1}-\textbf{U}\mathbf{\Sigma_{\alpha\alpha}}\textbf{V}^\hr)^{-\tr} \left(\rho_\textit{q}\h?{r-u}^\hr(\textbf{U}\mathbf{\Sigma_{\alpha\beta}}\textbf{V}^\hr)^\tr \right)^\tr \\
& \left(\rho_\textit{q}\h?{r-u}^\hr(\textbf{U}\mathbf{\Sigma_{\alpha\beta}}\textbf{V}^\hr)^\tr{(\mathbf{\Upsilon}_\textit{q}^{-1}-\textbf{U}\mathbf{\Sigma_{\alpha\alpha}}\textbf{V}^\hr)^{-1}}(\textbf{U}\mathbf{\Sigma_{\alpha\beta}}\textbf{V}^\hr)\h?{b-r}\f_\textit{q}-\I_K\right)^* \\
& \left((\textbf{U}\mathbf{\Sigma_{\alpha\beta}}\textbf{V}^\hr)\h?{b-r}\f_\textit{q}\right)^{\tr}(\mathbf{\Upsilon}_{\textit{q}}^{-1}-\textbf{U}\mathbf{\Sigma_{\alpha\alpha}}\textbf{V}^\hr)^{-\tr},
\end{split}
\tag{19}
\end{align}
with $\mathbf{\Upsilon}_\textit{q}$, $\f_\textit{q}$, and $\rho_\textit{q}$ indicating the optimum RIS phase configuration, BS precoding matrix, and scaling parameter values for each $q$-th channel sample. Similarly, the gradient of (\ref{eqn:ao-phase-mata}) with respect to $\mathbf{\Sigma_{\alpha\beta}}$ can be approximated as: 
 \begin{equation}\label{eq:G_ab}  
 G_{\mathbf{\Sigma_{\alpha\beta}}}\triangleq
\frac{1}{Q}\sum_{q=1}^Q \left[\pdv{\textit{f}(\mathbf{\Sigma_{\alpha\alpha}},\mathbf{\Sigma_{\alpha\beta}})}{\mathbf{\Sigma_{\alpha\beta}}}\right]_\textit{q}, \tag{20}
\end{equation}
where the gradient for each $q$-th channel sample is given by:
\begin{align}
\scriptsize
\begin{split}
& \left[\pdv{\textit{f}(\mathbf{\Sigma_{\alpha\alpha}},\mathbf{\Sigma_{\alpha\beta}})}{\mathbf{\Sigma_{\alpha\beta}}}\right]_\textit{q}= 
  2\left(\rho_\textit{q}\h?{r-u}^\hr(\textbf{V}^\hr)^\tr \right)^\tr \\
& \left(\rho_\textit{q}\h?{r-u}^\hr(\textbf{U}\mathbf{\Sigma_{\alpha\beta}}\textbf{V}^\hr)^\tr{(\mathbf{\Upsilon}_{\textit{q}}^{-1}-\textbf{U}\mathbf{\Sigma_{\alpha\alpha}}\textbf{V}^\hr)^{-1}}(\textbf{U}\mathbf{\Sigma_{\alpha\beta}}\textbf{V}^\hr)\h?{b-r}\f_{\textit{q}}-\I_K\right)^* \\
& (\textbf{V}^\hr\h?{b-r}\f_{\textit{q}})^{\tr}\mathbf{\Sigma_{\alpha\beta}}^\tr
\left(\textbf{U}^\tr(\mathbf{\Upsilon}_{\textit{q}}^{-1}-\textbf{U}\mathbf{\Sigma_{\alpha\alpha}}\textbf{V}^\hr)^{-1}\textbf{U}\right)^\tr + \\
& 2\left(\textbf{U}^\tr(\mathbf{\Upsilon}_{\textit{q}}^{-1}-\textbf{U}\mathbf{\Sigma_{\alpha\alpha}}\textbf{V}^\hr)^{-1}\textbf{U}\right)^\tr\mathbf{\Sigma_{\alpha\beta}}^\tr\left(\rho_\textit{q}\h?{r-u}^\hr(\textbf{V}^\hr)^\tr\right)^\tr \\ & \left(\rho_\textit{q}\h?{r-u}^\hr(\textbf{U}\mathbf{\Sigma_{\alpha\beta}}\textbf{V}^\hr)^\tr{(\mathbf{\Upsilon}_{\textit{q}}^{-1}-\textbf{U}\mathbf{\Sigma_{\alpha\alpha}}\textbf{V}^\hr)^{-1}}(\textbf{U}\mathbf{\Sigma_{\alpha\beta}}\textbf{V}^\hr)\h?{b-r}\f_{\textit{q}}-\I_K\right)^* \\
& (\textbf{V}^\hr\h?{b-r}\f_{\textit{q}})^{\tr}.
\end{split}
\tag{21}
\end{align}
Once the gradients are computed, the parameters $\mathbf{\Sigma_{\alpha\alpha}}$ and $\mathbf{\Sigma_{\alpha\beta}}$ are updated as follows:
\begin{align}
\mathbf{\Tilde{\Sigma}_{\alpha\alpha}}&=\mathbf{\Sigma_{\alpha\alpha}}-\mu{\rm Re}\left\{G_{\mathbf{\Sigma_{\alpha\alpha}}}\right\}, \tag{22}
\\
\mathbf{\Tilde{\Sigma}_{\alpha\beta}}&=\mathbf{\Sigma_{\alpha\beta}}-\mu{\rm Re}\left\{G_{\mathbf{\Sigma_{\alpha\beta}}}\right\}, \tag{23}
\end{align}
where we have adopted the real part of the gradients as the search directions and $\mu$ denotes the fixed step size. To maintain the symmetry of $\mathbf{\Tilde{\Sigma}_{\alpha\alpha}}$ as per (\ref{eqn:eq91-c}), we symmetrically update its diagonal elements at positions $\Tilde{\sigma}_{\alpha\alpha,ii}$ and $\Tilde{\sigma}_{\alpha\alpha,(M-i+1)(M-i+1)}$ $\forall$$i=1,2,\ldots,M$ by taking their average, as follows:
\begin{align}
\label{eq:symmetric}
&\Tilde{\sigma}_{\alpha\alpha,ii}=\frac{\Tilde{\sigma}_{\alpha\alpha,ii}+ \Tilde{\sigma}_{\alpha\alpha,(M-i+1)(M-i+1)}}{2} \tag{24},\\
\label{eq:symmetric_1}
&\Tilde{\sigma}_{\alpha\alpha,(M-i+1)(M-i+1)}=\frac{\Tilde{\sigma}_{\alpha\alpha,ii}+\Tilde{\sigma}_{\alpha\alpha,(M-i+1)(M-i+1)}}{2} \tag{25},
\end{align}
Clearly, when $\mathbf{\Sigma_{\alpha\alpha}}$ is symmetric, $\mathbf{\Sigma_{\alpha\beta}}$ becomes symmetric as well since $\mathbf{I}_M$ is also symmetric. Therefore, we also make $\mathbf{\Tilde{\Sigma}_{\alpha\beta}}$ symmetric using similar expressions to (\ref{eq:symmetric}) and (\ref{eq:symmetric_1}). 

We now formulate the following projection problem that considers both constraints \eqref{eqn:eq91-b} and \eqref{eqn:eq91-c} for the design of symmetric $\mathbf{\Sigma_{\alpha\alpha}}$ and $\mathbf{\Sigma_{\alpha\beta}}$:
\begin{subequations}
\label{eqn:pgg} 
\begin{align*}
   \tag{26a}
    \begin{split}
     \label{eqn:pg}
     \ds\argmin_{ \mathbf{\Sigma_{\alpha\alpha}},  \mathbf{\Sigma_{\alpha\beta}}} \quad & \norm{\mathbf{{\Sigma}}_{\alpha\alpha}-\mathbf{\Tilde{\Sigma}}_{\alpha\alpha}}?F^2 +\norm{\mathbf{\Sigma}_{\alpha\beta}-\mathbf{\Tilde{\Sigma}}_{\alpha\beta}}?F^2
    \end{split} \\
    \vspace{5pt}
\label{eqn:pg_1} \tag{26b} 
\tn{subject to} \quad &
\mathbf{\Sigma}_{\alpha\alpha}\mathbf{\Sigma}_{\alpha\alpha}^\hr+\mathbf{\Sigma}_{\alpha\beta}\mathbf{\Sigma}_{\alpha\beta}^\hr=\I_{M}. \end{align*}  
\end{subequations} 
We then simplify the term $\norm{\mathbf{\Sigma_{\alpha\alpha}}-\mathbf{\Tilde{\Sigma}_{\alpha\alpha}}}?F^2$ as follows:
\begin{align} \tag{27}
\small
\label{eqn:eq333}
\begin{split}
        &\norm{\mathbf{\Sigma_{\alpha\alpha}}-\mathbf{\Tilde{\Sigma}_{\alpha\alpha}}}?F^2 ~=~ \tn{Tr}\bigl((\mathbf{\Sigma_{\alpha\alpha}}-\mathbf{\Tilde{\Sigma}_{\alpha\alpha}})(\mathbf{\Sigma_{\alpha\alpha}}-\mathbf{\Tilde{\Sigma}_{\alpha\alpha}})^\hr\bigr)\\ &
        ~=~ \sum_{i=1}^{M} \sigma_{\alpha\alpha,ii}^2 -2\sigma_{\alpha\alpha,ii}\Tilde{\sigma}_{\alpha\alpha,ii}+\Tilde{\sigma}_{\alpha\alpha,ii}^2,
    \end{split}
\end{align}
\normalsize
where $\sigma_{\alpha\alpha,ii}$ and $\Tilde{\sigma}_{\alpha\alpha,ii}$ are the diagonal elements of $\mathbf{\Sigma_{\alpha\alpha}}$ and $\mathbf{\Tilde{\Sigma}_{\alpha\alpha}}$, respectively. Similarly, the other term $\norm{\mathbf{\Sigma_{\alpha\beta}}-\mathbf{\Tilde{\Sigma}_{\alpha\beta}}}?F^2$ in the objective can be simplified as:
\begin{align} \tag{28}
\small
\label{eqn:eq333_1}
\begin{split}
        &\norm{\mathbf{\Sigma_{\alpha\beta}}-\mathbf{\Tilde{\Sigma}_{\alpha\beta}}}?F^2 ~=~ \tn{Tr}\bigl((\mathbf{\Sigma_{\alpha\beta}}-\mathbf{\Tilde{\Sigma}_{\alpha\beta}})(\mathbf{\Sigma_{\alpha\beta}}-\mathbf{\Tilde{\Sigma}_{\alpha\beta}})^\hr\bigr)\\ &
        ~=~ \sum_{i=1}^{M} \sigma_{\alpha\beta,ii}^2 -2\sigma_{\alpha\beta,ii}\Tilde{\sigma}_{\alpha\beta,ii}+\Tilde{\sigma}_{\alpha\beta,ii}^2,
    \end{split}
\end{align}
\normalsize
where the scalars $\sigma_{\alpha\beta,ii}$ and $\Tilde{\sigma}_{\alpha\beta,ii}$ represent the diagonal elements of the matrices $\mathbf{\Sigma_{\alpha\beta}}$ and $\mathbf{\Tilde{\Sigma}_{\alpha\beta}}$, respectively. Using the latter expressions, the projection problem in \eqref{eqn:pg} and \eqref{eqn:pg_1} can be reformulated as follows:
\begin{subequations}
\label{eqn:eq_new_1}
\begin{align}
   \tag{29a}
    \begin{split}
     \label{eqn:new_1}
      \argmin_{ \sigma_{\alpha\alpha,ii},  \sigma_{\alpha\beta,ii}} \,\, &\sum_{i=1}^{M} (\sigma_{\alpha\alpha,ii}^2 -2\sigma_{\alpha\alpha,ii}\Tilde{\sigma}_{\alpha\alpha,ii}+\Tilde{\sigma}_{\alpha\alpha,ii}^2) \\ &+ \sum_{i=1}^{M} (\sigma_{\alpha\beta,ii}^2 -2\sigma_{\alpha\beta,ii}\Tilde{\sigma}_{\alpha\beta,ii}+\Tilde{\sigma}_{\alpha\beta,ii}^2)
    \end{split} \\
\label{eqn:new_2} \tag{29b} 
\tn{subject to} \,\,
& \sigma_{\alpha\alpha,ii}^2+\sigma_{\alpha\beta,ii}^2=1,  \,\, \forall i=1,2,\ldots,M.
\end{align}  
\end{subequations} 

In the sequel, we present closed-form solutions for the latter problem using the method of Lagrange multipliers.

\subsubsection{Closed-Form Solution}
The Lagrangian function of the latter optimization problem is defined as follows:
\begin{align} 
 &\nonumber L(\sigma_{\alpha\alpha,ii},\sigma_{\alpha\beta,ii},\lambda_i)\triangleq\sum_{i=1}^{M} (\sigma_{\alpha\alpha,ii}^2 -2\sigma_{\alpha\alpha,ii}\Tilde{\sigma}_{\alpha\alpha,ii}+\Tilde{\sigma}_{\alpha\alpha,ii}^2) \nonumber \\ &  +\sum_{i=1}^{M} (\sigma_{\alpha\beta,ii}^2 -2\sigma_{\alpha\beta,ii}\Tilde{\sigma}_{\alpha\beta,ii}+\Tilde{\sigma}_{\alpha\beta,ii}^2)\nonumber  \\ & +\sum_{i=1}^{M}\lambda_i(\sigma_{\alpha\alpha,ii}^2 +\sigma_{\alpha\beta,ii}^2 -1).\tag{30}\nonumber 
\end{align}
\normalsize
Its partial derivatives with respect to the optimization variables $\sigma_{\alpha\alpha,ii}$, $\sigma_{\alpha\beta,ii}$, and $\lambda_i$ are then computed as:
\begin{align} \tag{31}
\label{eqn:eq333_1_1}
&\pdv{L(\sigma_{\alpha\alpha,ii},\sigma_{\alpha\beta,ii},\lambda_i)}{\sigma_{\alpha\alpha,ii}}=
2\sigma_{\alpha\alpha,ii}-2\Tilde{\sigma}_{\alpha\alpha,ii}+2\lambda_i\sigma_{\alpha\alpha,ii},
\end{align}
\normalsize
\begin{align} \tag{32}
\label{eqn:eq333_1_2}
&\pdv{L(\sigma_{\alpha\alpha,ii},\sigma_{\alpha\beta,ii},\lambda_i)}{\sigma_{\alpha\beta,ii}}= 2\sigma_{\alpha\beta,ii}-2\Tilde{\sigma}_{\alpha\beta,ii}+2\lambda_i\sigma_{\alpha\beta,ii},
\end{align}
\normalsize
\begin{align} \tag{33}
\label{eqn:eq333_1_3}
&\pdv{L(\sigma_{\alpha\alpha,ii},\sigma_{\alpha\beta,ii},\lambda_i)}{\lambda_i} = \sigma_{\alpha\alpha,ii}^2+\sigma_{\alpha\beta,ii}^2-1. 
\end{align}
\normalsize
By equating each of the latter expressions to zero and then solving the resulting $3\times3$ system of equations, the diagonal elements of $\mathbf{\Sigma_{\alpha\alpha}}$ and $\mathbf{\Sigma_{\alpha\beta}}$ are obtained in closed-form as:
\begin{equation} \tag{34}
\label{eq:sol1}
\sigma_{\alpha\alpha,ii}=\frac{\Tilde{\sigma}_{\alpha\alpha,ii}}{\sqrt{\Tilde{\sigma}_{\alpha\alpha,ii}^2+\Tilde{\sigma}_{\alpha\beta,ii}^2}},
\end{equation}
\begin{equation} \tag{35}
\label{eq:sol1}
\sigma_{\alpha\beta,ii}=\frac{\Tilde{\sigma}_{\alpha\beta,ii}}{\sqrt{\Tilde{\sigma}_{\alpha\alpha,ii}^2+\Tilde{\sigma}_{\alpha\beta,ii}^2}}.
\end{equation}
The solution of the overall system design problem is summarized in \textbf{Algorithm} \ref{algo:joint_op}.
\begin{algorithm}
\caption{Proposed System Design}
\label{algo:joint_op}
\begin{algorithmic}[1]
\renewcommand{\algorithmicrequire}{\textbf{Input:}}
\renewcommand{\algorithmicensure}{\textbf{Output:}}
\REQUIRE $Q$ samples of the channel matrices $\h?{b-r}$ and $\h?{r-u}$.
\STATE Set $\textbf{U}=\textbf{V}=\mathbf{D} \otimes\mathbf{D}$.
\vspace{1pt}
\STATE Initialize $\widehat{\mathbf{\Sigma}}_{\alpha\alpha_{0}}$ and $\widehat{\mathbf{\Sigma}}_{\alpha\beta_{0}}$. 
\FOR{$k=1,2,\ldots,I_{\max}$}
\FOR{$q=1,2,\ldots,Q$}
\STATE Initialize $\mathbf{\Upsilon}_k$.
\STATE Initialize $\rho_k$ and $\f_k$ as described in \cite{10096563}.
\STATE Compute $\mathbf{\Upsilon}_k$ and $\f_k$ solving sub-problem 1) via the method in \cite{10096563}.
\STATE Compute $\left[\pdv{\textit{f}(\mathbf{\Sigma}_{\alpha\alpha,k},\mathbf{\Sigma}_{\alpha\beta,k})}{\mathbf{\Sigma}_{\alpha\alpha,k}}\right]_\textit{q}$.
\STATE Compute $\left[\pdv{\textit{f}(\mathbf{\Sigma}_{\alpha\alpha,k},\mathbf{\Sigma}_{\alpha\beta,k})}{\mathbf{\Sigma}_{\alpha\beta,k}}\right]_{\textit{q}}$.
\ENDFOR
\STATE Compute $G_{\mathbf{\Sigma}_{\alpha\beta,k}}$ using \eqref{eq:G_aa}.  
\STATE Compute $G_{\mathbf{\Sigma}_{\alpha\beta,k}}$ using \eqref{eq:G_ab}.
\STATE Compute $\mathbf{\Sigma}_{\alpha\alpha,k}=\mathbf{\Sigma}_{\alpha\alpha,k-1}-\mu{\rm Re}\left\{G_{\mathbf{\Sigma}_{\alpha\alpha,k}}\right\}$.
\STATE Compute 
$\mathbf{\Sigma}_{\alpha\beta,k}=\mathbf{\Sigma}_{\alpha\beta,k-1}-\mu{\rm Re}\left\{G_{\mathbf{\Sigma}_{\alpha\beta,k}}\right\}$.
\vspace{1.5pt}
\STATE Use (\ref{eq:symmetric}) and (\ref{eq:symmetric_1}) to symmetrify $\mathbf{\Sigma}_{\alpha\alpha,k}$ and $\mathbf{\Sigma}_{\alpha\beta,k}$.
\vspace{1.5pt}
\STATE Compute $\widehat{\sigma}_{\alpha\alpha,ii,k}=\frac{\sigma_{\alpha\alpha,ii,k}}{\sqrt{\sigma_{\alpha\alpha,ii,k}^2+\sigma_{\alpha\beta,ii,k}^2}}$.
\vspace{1.5pt}
\STATE Compute $\widehat{\sigma}_{\alpha\beta,ii,k}=\frac{\sigma_{\alpha\beta,ii,k}}{\sqrt{\sigma_{\alpha\alpha,ii,k}^2+\sigma_{\alpha\beta,ii,k}^2}}$.
\vspace{1.5pt}
\STATE Set $\mathbf{S}_{\alpha\alpha,k}=\textbf{U}\tn{Diag}\left(\left[\widehat{\sigma}_{\alpha\alpha,11,k},\ldots,\widehat{\sigma}_{\alpha\alpha,MM,k}\right]\right)\textbf{V}^\hr$.
\STATE Set $\mathbf{S}_{\alpha\beta,k}=\textbf{U}\tn{Diag}\left(\left[\widehat{\sigma}_{\alpha\beta,11,k},\ldots,\widehat{\sigma}_{\alpha\beta,MM,k}\right]\right)\textbf{V}^\hr$.
\ENDFOR
\ENSURE $\rho_{I_{\max}}$, $\f_{I_{\max}}$, $\mathbf{\Upsilon}_{I_{\max}}$, $\mathbf{S}_{\alpha\alpha,I_{\max}}$, and $\mathbf{S}_{\alpha\beta,I_{\max}}$.
\end{algorithmic}
\end{algorithm}

\begin{figure*}
\centering
\begin{subfigure}{.333\textwidth}
\centering
\includegraphics[scale=0.27]{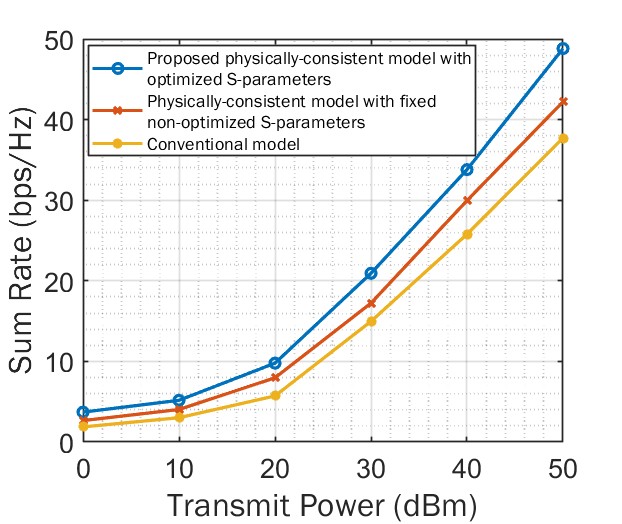}
\caption{\footnotesize $Q=10$, $K=6$, $N=32$, and $M=64$.}
\label{fig:sumrate_64_1}
\end{subfigure}%
\begin{subfigure}{.334\textwidth}
\centering
\includegraphics[scale=0.27]{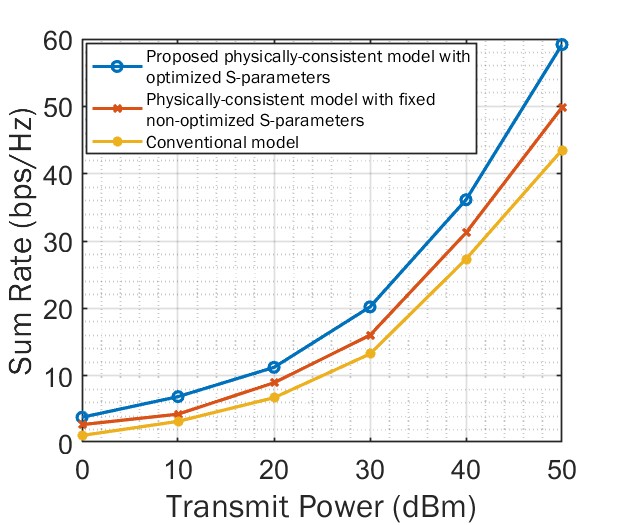}
\caption{\footnotesize $Q=10$, $K=8$, $N=32$, and $M=64$.}
\label{fig:sumrate_64_2}
\end{subfigure}%
\begin{subfigure}{.333\textwidth}
\centering
\includegraphics[scale=0.27]{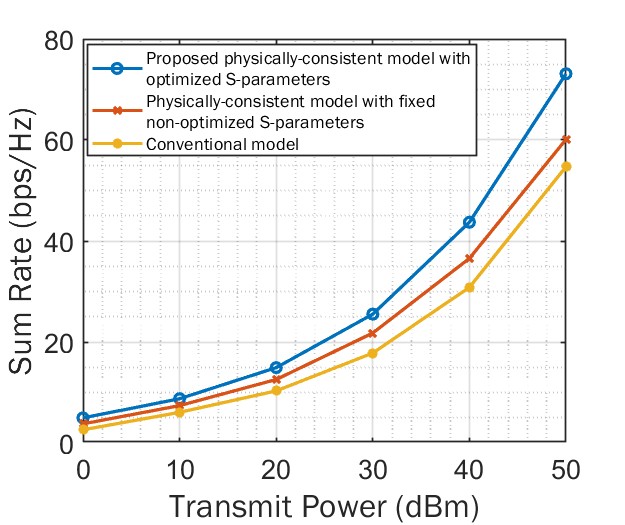}
\caption{\footnotesize $Q=6$, $K=8$, $N=32$, and $M=100$.}
\label{fig:sumrate_64_3}
\end{subfigure}
\caption{Sum-rate performance versus the transmit power $P$ in dBm for various system parameters.}
\label{fig:sumrate_64}
\end{figure*}
\begin{figure}[h]
\centering
\includegraphics[scale=0.29]{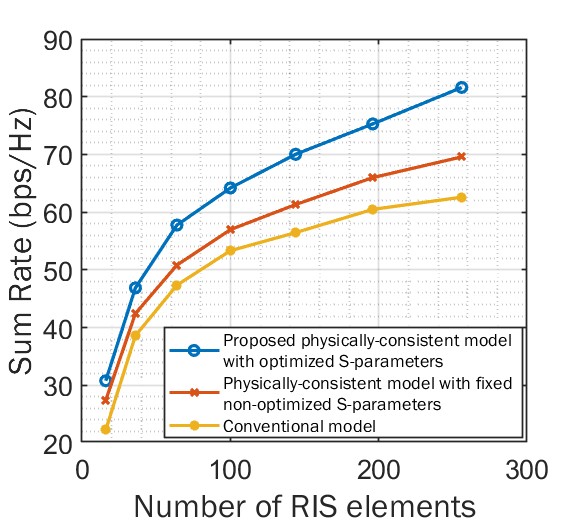}
\caption{Sum-rate performance versus the number of RIS elements $M$ considering $Q=10$ channel samples, $K=5$, $P=50$ dBm, and $N=32$.}
\label{fig:4}
\end{figure}

\section{Numerical Results}
In this section, we present computer simulation results for the proposed physically consistent RIS-aided multi-user MIMO design. We have considered the same simulation parameters with \cite{10096563} as well as the same parametric modeling for the channel matrices $\h?{b-r}$ and $\h?{r-u}$. For benchmarking, we have considered the following two baseline schemes:
\begin{itemize}
    \item The physically consistent design of \cite{10096563} that incorporates MC, but considers a fixed non-optimized scattering matrix $\mathbf{S_{\alpha\alpha}}$ and $\mathbf{S_{\alpha\beta}}=\mathbf{I}_M$.
    \item The scheme presented in \cite{9474428} that is based on a conventional (hypothetical) model that does not consider the presence of MC.
\end{itemize}
It can be observed in Fig.~\ref{fig:sumrate_64} that there is an improvement in the sum rate when optimizing MC via the proposed offline approach, as compared to the other two benchmarks. In addition, Fig.~\ref{fig:4} illustrates the variation of the sum rate versus the number $M$ of RIS elements, and a considerable improvement is evident with the proposed optimization approach. Interestingly, to satisfy a given desired surface area, the RIS elements need to be placed in closer proximity as $M$ increases. However, placing elements closer results in stronger MC, implying that an increase in $M$ leads to higher sum-rate performance when optimizing the MC. It can be thus concluded from all simulation results that it suffices for the MC to be optimized offline to gain from its optimization. 

\section{Conclusion}
In this paper, we presented a novel framework to optimize the RIS phase configuration and BS active precoders using a physically-consistent end-to-end channel model incorporating MC. We proposed an offline optimization method, applied to a class of wireless channels, to design MC at the RIS, as opposed to state-of-the-art approaches that did not perform MC optimization. The resultant nested optimization problem has been decomposed into two sub-problems. In addition, the RIS phase configuration and BS active precoding matrices were designed using online optimization. Our approach emphasizes the potential of ``engineering" MC to enhance sum-rate performance without requiring on-the-fly adjustments. By incorporating optimized MC into active and passive beamforming, the proposed framework provides enhanced system performance.

\bibliographystyle{IEEEtran}
\bibliography{bib}
\end{document}